\documentclass[aps,pre,twocolumn,floatfix,
nofootinbib,showpacs,longbibliography]{revtex4-1}

\usepackage{mathtools}
\usepackage{color}
\usepackage{color}
\usepackage{blkarray, bigstrut}
\usepackage{relsize}
\usepackage{amsmath}
\usepackage[british]{babel}  
\usepackage[scaled=1.03]{inconsolata} 

\usepackage[colorlinks=true, citecolor=blue, urlcolor=blue]{hyperref}  
\usepackage{graphicx} 
\usepackage[babel]{microtype}  
\usepackage{amsmath,amssymb,amsthm,bm,amsfonts,mathrsfs,bbm} 
\usepackage{xspace}  
\usepackage{pgfplots}
\usepackage{amsmath}
\usepackage{amssymb}

\begin{document}

\title{Bound on Ergotropic Gap for Bipartite Separable States}

\author{Mir Alimuddin}
\author{Tamal Guha}
\author{Preeti Parashar}
\affiliation{Physics and Applied Mathematics Unit, Indian Statistical Institute, 203 B T Road, Kolkata-700108, India.}
 

\begin{abstract} 

Presence of correlations among the constituent quantum systems has a great relevance in thermodynamics. Significant efforts have been devoted to investigate the role of correlations in work extraction, among others. Here, we derive a bound on the difference between global and local extractable work by unitary operations (ergotropic gap), for bipartite separable states. Violation of this bound necessarily certifies  the presence of entanglement. This gap is shown to be a monotone under LOCC assisted state transformations for pure bipartite quantum states. Our criterion has an implication in witnessing the dimension of a bipartite quantum state, with same local dimensions. On the other hand, our result gives an operational meaning to the Nielsen-Kempe disorder criterion. We also propose a schematic model to realize the separability bound experimentally and to detect entanglement for a restricted class of quantum states.
\end{abstract}  
\maketitle

\section{Introduction} \label{sec1} Correlations among various systems are defined in terms of several information theoretic quantities, viz., mutual information, conditional entropy, accessible information etc. \cite{NielsenBook, wildebook,Statemerging, Holevo'73}. Quantum theory provides a larger class of possible correlations and it is interesting to study whether these quantum correlations provide some dramatic supremacy over the classical regime.

Although, the advantage of quantum entanglement is well explored in numerous information theoretic and computational tasks \cite{densecoding,teleportation,ekart,rac,bayesian}, the power of this correlation has recently been studied in various work extraction protocols from the thermodynamic perspective \cite{Oppenheim'02,Oppenheim'PRL,allahavardyan,Vedral'05,alicki,huberPRX,manikPRE,Ciampini,Bera'17,Francica'npj}. This motivates us to specify the following task:  Consider a bipartite quantum system  governed by the local linear spaced Hamiltonian. The parties can use local unitary (generated from a cyclic potential alongside their local Hamiltonians) on their individual quantum system to extract the maximum amount of work called local ergotropy \cite{alicki, allahavardyan}. Now if these two parties are allowed to come together and collaborate, can they extract more work jointly?
This difference between the global and total local ergotropy of quantum systems is called the ergotropic gap \cite{huberPRX, manikPRE}.

In this paper, we present a bound on the ergotropic gap for bipartite separable states of arbitrary dimension. Any value beyond this bound necessarily implies the presence of entanglement and gives {\it quantum advantage}. Thus our result gives an {\it operational thermodynamic criterion} for entanglement detection.
Although we derive a computable bound on ergotropic gap, this criterion becomes necessary and sufficient for the class of states with maximally mixed marginals. 
The criterion is experimentally realizable for a larger class of states whose marginals are passive in nature. We propose a simple model to detect entanglement based on our bound by implementing global unitary operations.

In \cite{huberPRX}, the authors have shown that among all possible correlated states with constant marginal entropy, the ergotropic gap for entangled states is maximum.  Here we establish a relation  between entanglement and ergotropic gap for pure bipartite states;  the one with more entanglement will give higher ergotropic gap. For the case of two-qubit systems, the converse also holds.

Witnessing the dimension of a
given state depends on several statistical criteria and is a
subject of recent interest \cite{Brunner'08,Brunner'10,Brunner'13,Arup'PRA}.
In the present work, we show that the ergotropic gap can be used as a dimension witness, which gives a lower bound on the dimension of a $d\times d$ bipartite state.

The article is organized as follows: In Section II we discuss the framework of work extraction and the effect of various types of correlations on the ergotropic gap. Section III contains our main results on the bound on ergotropic gap for separable states. We also compare the ergotropic gap with entanglement for pure states. In Section IV, an operational interpretation of the bound is given as a thermodynamic criterion for separability, and its experimental implementation is outlined in Section VI. The extractable work difference between bath assisted global and local systems is shown to be related to quantum mutual information in Section V.
We show that the ergotropic gap behaves like a dimension witness in Section VII and finally the conclusions in Section VIII. The detailed proof of Theorem 2 is worked out in the Appendix.

\section{Framework}
\subsection{ Work Extraction}
One of the most important operational quantities in thermodynamics is work. There are mainly two different approaches to extract work from the given system: either by using the system along with a thermal bath and applying the  global unitary on the joint system \cite{popescuNAT}, or a cyclic Hamiltonian $H(t)=H+V(t)$ can be applied to the system, i.e., the state is evolved under a unitary $U(\tau)= \overrightarrow\exp\left(-i\hbar\int_0^\tau d t\left(H + V(t)\right)\right)$ commuting with the total Hamiltonian. Here, the time-dependent potential $V(t)$ starts at $t=0$ and decouples from the system at $t=\tau$ \cite{allahavardyan,alicki}.

Under this action of the unitary, the initial quantum state $\rho_{i}$ evolves to $\rho_{f}$. The final state $\rho_{f}$ will be such that no work can be extracted from it (single copy) under the unitary action. Such a state is defined as a passive state . It necessarily implies that the state will be (block) diagonal in the Hamiltonian basis and the population will be inverse to the increasing energy states. But the scope remains open for more copies of the passive state $\rho_{f}$. The state is said to be completely passive or thermal in nature if {\it no work} can be extracted even if infinite copies of $\rho_{f}$ are used jointly with full access to the global unitary  \cite{lenard,skrzypzyckPRE,Pusz'CMP}.

\subsection{Correlation and Ergotropic gap}
While accessing the global state during work extraction, one can exploit the correlations present in it. So it is natural to  ask whether \textit{quantum} correlations give more global advantage than \textit{classical} correlations. Recently, this has been  answered affirmatively in \cite{huberPRX}, by considering identical local marginals. However, it is difficult to characterize the explicit connection between correlations and ergotropy gap in the general scenario. In this article, we are going to investigate this connection for general bipartite systems.

  Let us consider a bipartite state $\rho^{AB}\in\mathscr{D}(\mathscr{H}_A\otimes \mathscr{H}_B)$ where, $\mathscr{H}_{A/B}$ corresponds to the Hilbert space of $A/B$ subsystem and $\mathscr{D}(\mathscr{H}_A\otimes \mathscr{H}_B)$ refers to the set of density operators of the composite state. The $i^{th}$ subsystem is governed by the Hamiltonian $H_i= \sum_{j}{{\epsilon_{j}^{i}}^{\uparrow}}  |j\rangle^{i}\langle j| $, where ${{\epsilon_{j}^{i}}^{\uparrow}}$ and $|j\rangle^{i}$ is the $j$th energy eigenvalue and energy eigenvector for $i^{th}$ Hamiltonian. In the case of linear Hamiltonian,  ${\epsilon^{i}_{j}} = j {\epsilon^{i}}$. 
   The total interaction free global Hamiltonian is $H_g= H_A \otimes I_B + I_A \otimes  H_B$. 
  In this premise, maximum work is extracted from the isolated bipartite state $\rho^{AB}$ by transforming it to the corresponding passive state ${\rho^{AB}_{p}}$ under a cyclic unitary operator $U(\tau)$, where $U(\tau)$ is controlled by the external potential $V(t)$ acting cyclically over the time interval $0 \leq t \leq \tau$ on the global system. The maximum extractable work termed as ergotropy is defined by 
  \begin{equation}
  \begin{aligned}
  \mathcal{W}^{g}_e &= Tr(\rho^{AB} H_g) - \min_{U\in \mathscr{L}(\mathscr{H}_A\otimes \mathscr{H}_B)}Tr\{U\rho^{AB}U^{\dagger} H_g\} \\
  &= Tr(\rho^{AB} {H}_g) - Tr(\rho^{AB}_{p}{H}_g)\label{globalergo}, 
  \end{aligned}
  \end{equation}
 where $\mathscr{L}(X)$ denotes the set of all bounded linear operators on the Hilbert space $X$.
  
  It may happen that not the whole system but rather its reduced parts are accessible to Alice and Bob individually. In this case they choose some proper potential $V_{i}(t)$ for the time interval $0 \leq t \leq \tau$ corresponding to their individual unitary operator $U_{A/B}(\tau)$. The total achievable work called local ergotropy, is defined by
  \begin{equation}
  \mathcal{W}^{l}_e = \mathcal{W}^{A}_e + \mathcal{W}^{B}_e ,   
  \end{equation}
   where $\mathcal{W}^{A}_e$ and $\mathcal{W}^{B}_e$ is the maximum extractable work in Alice and Bob's lab respectively, written as, 
   	\begin{eqnarray}
   	\begin{aligned}
 \mathcal{W}^{A}_e & =  Tr(\rho^{AB}H_A \otimes I_B)\\
  & - \min_{U_A\in \mathscr{L}(\mathscr{H}_A)}Tr\{(U_A \otimes {I}_B) \rho^{AB} ({U_A} \otimes {I}_B)^{\dagger}(H_A \otimes I_B)\}
  \end{aligned}
  \end{eqnarray}
  and
  \begin{eqnarray}
  \begin{aligned}
 \mathcal{W}^{B}_e &= Tr(\rho^{AB} I_A\otimes H_B )\\ 
 &-\min_{U_{B}\in \mathscr{L}(\mathscr{H}_B)}Tr\{({I}_A\otimes U_B ) \rho^{AB} ({I}_A \otimes {U_B})^{\dagger} I_A \otimes H_B\}
 \end{aligned}
 \end{eqnarray}
 It follows that, 
 \begin{equation}
\mathcal{W}^{l}_e = Tr(\rho^{AB} H_g)  -\{ Tr(\rho^{A}_{p}H_A)+ Tr(\rho^{B}_{p}H_B)\},
 \end{equation}
 where $\rho^{A}_{p}$ and $\rho^{B}_{p}$ are the passive states for system $A$ and $B$ respectively.

 Now, we are in a position to define the advantage in terms of extractable work from global accessibility over the local one. This extra gain is called {\it ergotropy gap} defined by,

\begin{eqnarray}
\begin{aligned}
\Delta_{EG} &= \mathcal{W}^{g}_e - \mathcal{W}^{l}_e \\
 &= \{ Tr(\rho^{A}_{p}H_A)+ Tr(\rho^{B}_{p}H_B)\} - Tr(\rho^{AB}_{p} H_g) \label{eq:7}
\end{aligned}
\end{eqnarray}
 
  Global ergotropy is always greater or equal to the local one, as $ U_{A} \otimes U_{B}\subseteq U_{AB}$. Therefore, the ergotropic gap quantifies how much benefit can be obtained by doing global operations on the joint system instead of local operations. So clearly ergotropic gap depends on various kinds of correlations present in a bipartite quantum state as mentioned below.

Ergotropic gap for pure product states vanish, due to the fact that any pure state can be transformed to $|0\rangle$ under unitary operation. However, in general for product states the situation is different, depending upon individual Hamiltonian \cite{manikPRE}, where the ergotropic gap can be non-zero, even with vanishing local ergotropy. It is obvious that for several entangled or classically correlated states this gap is non-vanishing. However, in contrast to \cite{manikPRE} there is a class of entangled states where the ergotropic gap is washed out dramatically. This counter intuitive feature happens for identical local Hamiltonian, due to the existence of entangled states in the degenerate energy subspace.

 To get a flavour of the last statement let us consider the state $p |00\rangle \langle 00| + (1-p) | \psi^-\rangle\langle \psi^-|$. Due to the presence of $| \psi^-\rangle$ in the degenerate energy subspace spanned by $\{|01\rangle, |10\rangle\}$ the state remains passive globally when $p\ge \frac{1}{2}$, hence produces zero ergotropic gap. However the state is entangled, $\forall p \in [0,1)$.
    
So we see that correlation and ergotropy gap $\Delta_{EG}$ have a somewhat bizarre relation. Also the states that give quantum advantage remain uncharacterised. In the present work we give an optimal bound on ergotropic gap for all separable bipartite states for the above mentioned task. Ergotropic gap greater than this optimal value implies quantum advantage reflecting supremacy of quantum entanglement. The bound is derived as an implication of the Nielsen-Kempe disorder criterion \cite{nielsonPRL} which is summarized below.
   \begin{figure}[t!]
            \centering
            	\includegraphics[height=3cm,width=6cm]{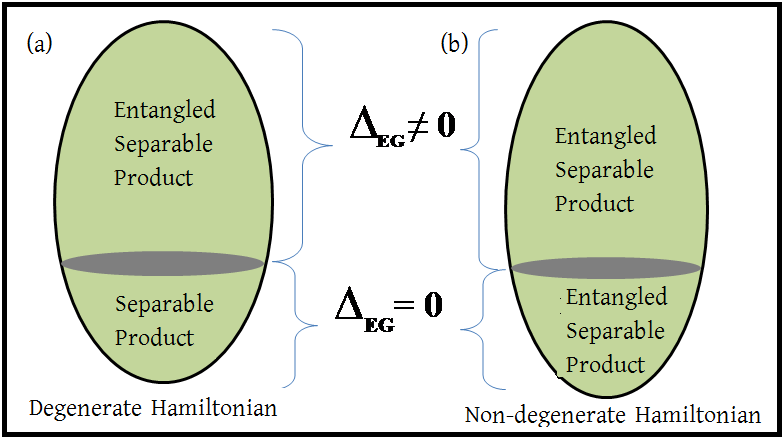}
            	\caption{(Color on-line) In (a) on the basis of $\Delta_{EG}$ we separate out the multipartite state space for non-degenerate Hamiltonian where entanglement certifies non zero ergotropic gap. In (b) we have shown the same separation but for degenerate Hamiltonian, where we get some entangled states with zero ergotropic gap.}\label{fig}
            \end{figure} 
 \subsection{Majorization Criterion}
 {\it Definition:} A state $\rho$ is said to be majorized by a state $\sigma$ i.e. $\lambda(\rho) \prec \lambda(\sigma)$ if,
 \begin{equation}
 \sum\limits_{i=1}^{k}p_{i}^{\downarrow} \leq \sum\limits_{i=1}^{k}q_{i}^{\downarrow} ~~~~ (1 \leq k \leq n-1)
 \end{equation}
  and
 \begin{equation}
  \sum\limits_{i=1}^{n}p_{i}^{\downarrow} = \sum\limits_{i=1}^{n}q_{i}^{\downarrow},
 \end{equation}
 where $\lambda(\rho) \equiv \{p_{i}^{\downarrow}\} \in \mathcal{R}^{n} $ and $\lambda(\sigma) \equiv  \{q_{i}^{\downarrow}\} \in \mathcal{R}^{n}$ are the spectrum of $\rho$ and $\sigma$ respectively, arranged in non-increasing order ($p_1^{\downarrow} \geq p_2^{\downarrow} \geq .~.~.~\geq p_n^{\downarrow}$), ($q_1^{\downarrow} \geq q_2^{\downarrow} \geq .~.~.~\geq q_n^{\downarrow}$).

  For different dimensions, extra zeros are appended to make the condition complete. Majorization criterion have great implication in state transformation in various resource theories\cite{Nielsen'99,Winter'PRL,Ng'18}. If $\rho \prec \sigma$ then it implies that $S(\rho) \geq S(\sigma)$ (but not the reverse) and  $\sigma \rightarrow \rho$ transition is possible under noisy evolution\cite{horodecki'03}.
   The notion of majorization was extended to give the following criteria of separability of a bipartite quantum state.
  
   {\it Nielsen-Kempe  disorder criterion of separability:} If $\rho^{AB}$ is separable, then
   \begin{equation}\label{NK}
   \lambda(\rho^{AB}) \prec \lambda(\rho^A)~~~ and~~~ \lambda(\rho^{AB}) \prec \lambda(\rho^B) ,
    \end{equation}
    where $\rho^A$ and $\rho^B$ are the states of system $A$ and $B$ respectively. It says that if the global state is separable, then it is more disordered than the local states \cite{nielsonPRL}.
  The above criterion is necessary for separability but not sufficient as the converse does not always hold. Although Eq.(\ref{NK}) is stronger than the entropic criterion for separability \cite{Wehrl'78}, it is weaker than the Reduction criterion\cite{Hiroshima'03}. This means that all the PPT \cite{Peres'96,Horodecki'96} and single copy non distillable states (Reduction criterion)\cite{Cerf'99,Horodecki'99,Breuer'06,Hall'06} satisfy the Nielsen-Kempe criterion but the reverse is not true. Since this criterion is spectral dependent, we use it to derive bounds on the ergotropy gap, which in turn provides an interesting physical interpretation of this criterion.
  
  \section{Bounds on ergotropic gap}
  
  A state with non zero ergotropic gap gives more work globally than locally. Although several product states can give non zero ergotropic gap, two-qubit product states (with identical local Hamiltonian) yield zero gap. The presence of correlation makes those states globally less disordered and as a consequence sometimes one is able to achieve higher ergotropic gap.  But, even in the case of the strongest correlation i.e., entanglement, there exist some states which give the same local and global ergotropy. However, entanglement is necessary to get {\it quantum advantage} and the maximum ergotropic gap is provided by  the maximally entangled states. Thus it is important to characterize the entangled states which give quantum advantage in ergotropic gap over all separable states. \\
 
 {\it {\bf Proposition:} A multipartite pure state governed by the general Hamiltonian is entangled if and only if it has non-zero ergotropic gap.} \\
 
 {\it {\bf Proof:}}  For pure product states local unitaries are sufficient to extract the maximum work. The initial state reaches the lowest energetic passive state $|0 \rangle ^{\otimes n}$ both locally as well as globally and thus $\Delta_{EG}$ becomes zero. For entangled states, the entangling unitary takes $|\psi\rangle^{ent}$ to $|0 \rangle ^{\otimes n}$ whereas local states are mixed and their local unitaries transform them to the minimum energetic but equal entropic passive states. So locally the accessible work becomes less which makes $\Delta_{EG} > 0$. 
 $~~~~~~~~~~~~~\blacksquare$\\

 {\it {\bf Theorem 1:}
 Ergotropic gap of a pure bipartite state $|\phi\rangle^{AB}$ is greater or equal to that of $|\psi\rangle^{AB}$ if $\lambda (|\phi\rangle)  \prec \lambda (|\psi\rangle) $, where $\lambda (|\phi\rangle)$ and $\lambda (|\psi\rangle)$  correspond to the spectrum of the individual marginals.}\\
  
 {\it {\bf Proof:}} 
 Consider two bipartite pure states in Schmidt form
 \begin{eqnarray}
 |\phi\rangle^{AB} = \sum\limits_{i=0}^{d_1-1} \sqrt{\lambda_i}|\alpha^A_i\rangle |\beta^B_i\rangle \nonumber\\ |\psi\rangle^{AB} = \sum\limits_{i=0}^{d_2-1} \sqrt{{\eta_i}}|a^A_i\rangle |b^B_i\rangle.\nonumber
  \end{eqnarray}
Here we assume that $d_1 \geq d_2$, i.e. $|\phi\rangle$ has more number of Schmidt coefficients than $|\psi\rangle$ and $\lambda_i$, $\eta_i$ have been chosen in non-increasing order.

  From any pure bipartite state it is always possible to reach the passive form $|00\rangle$ by some proper global unitary. The Schmidt decomposition gives the same spectrum for the marginals which can be written in the passive form in the energy basis as follows: 
  \begin{eqnarray}
 \rho^A_p(\phi)=\rho^B_p(\phi)= \sum\limits_{j=0}^{d_1-1}\lambda_j |j\rangle\langle j| \nonumber\\ 
 \rho^A_p(\psi)=\rho^B_p(\psi)= \sum\limits_{j=0}^{d_2-1}\eta_j |j\rangle\langle j|
 \end{eqnarray}
 
 The reduced systems $A$ and $B$ are governed by the Hamiltonian ${H}_A= \sum_{j}{\epsilon_{j}}^{A}  |j\rangle\langle j| $ and  ${H}_B= \sum_{j}{\epsilon_{j}}^{B}  |j\rangle\langle j| $ respectively.
  According to Eq. (\ref{eq:7}) ergotropic gap for state $|\phi\rangle$ and $|\psi\rangle$ would be
   \begin{eqnarray}
   \begin{aligned}
  \Delta_{EG}(\phi)
  &= \sum\limits_{j=0}^{d_1-1}\lambda_j \epsilon^A_j + \sum\limits_{j=0}^{d_1-1}\lambda_j \epsilon^B_j - Tr(|00\rangle\langle 00| {H}_g) \\ 
  &=\sum\limits_{j=0}^{d_1-1}\lambda_j\epsilon^{AB}_j-Tr(|00\rangle\langle 00| {H}_g)\\
  \end{aligned}
   \end{eqnarray}
   and
  \begin{eqnarray}
  \Delta_{EG}(\psi) = \sum\limits_{j=0}^{d_2-1}\eta_j\epsilon^{AB}_j-Tr(|00\rangle\langle 00| {H}_g),
  \end{eqnarray}
    where $\epsilon^{AB}_j = \epsilon^{A}_j +\epsilon^{B}_j $ is the energy for the corresponding $|jj\rangle$ state.\\
    
    The difference between the two ergotropic gaps
     is
    \begin{eqnarray}
    \Delta_{EG}(\phi)-\Delta_{EG}(\psi) = \sum\limits_{j=d_2}^{d_1-1}\lambda_j\epsilon^{AB}_j+ \sum\limits_{j=0}^{d_2-1}(\lambda_j-\eta_j)\epsilon_j^{AB} \nonumber\\
    = \sum\limits_{j=d_2}^{d_1-1}\lambda_j\epsilon^{AB}_j+ \sum\limits_{k=0}^{d_2-2}(\epsilon^{AB}_{k+1}-\epsilon_k^{AB})\sum\limits_{j=0}^{k}(\eta_j-\lambda_j).\nonumber\\
    \end{eqnarray}
    
 If the majorization condition holds i.e. $\sum\limits_{j=0}^{k} \lambda_i \leq \sum\limits_{j=0}^{k} \eta_i$, for any $k \geq 0$,   then $\Delta_{EG}(\phi)-\Delta_{EG}(\psi)\geq 0$. $~~~\blacksquare$ \\
 
 {\it {\bf Corollary 1.1:} For the case of pure two-qubit system, $\Delta_{EG}$ becomes an entanglement measure which is robust in nature.}
 
 {\it Proof:}
 Consider a pure two-qubit state
  \begin{eqnarray}
  |\chi\rangle^{AB} = \sqrt{{\lambda_{max}}} {| \psi \rangle}^A \otimes {| \phi \rangle}^B +  \sqrt{{\lambda_{min}}} {| \psi \rangle^\perp}^A \otimes {| \phi \rangle^\perp}^B \nonumber\\
  \end{eqnarray}
  with the marginals governed by the Hamiltonian ${H_{A}}=E_a |1\rangle \langle1|$ and ${H_{B}}=E_b |1\rangle \langle1|$ and the global Hamiltonian is ${H}_g= E_a|10\rangle\langle 10| + E_b |01\rangle \langle 01| + (E_a+E_b) |11\rangle \langle 11|$, where the ground state energy for individual systems are scaled to zero.
  \par
  By some proper global unitary we can transfer $|\chi\rangle$ to its passive state $|00\rangle$ to extract global ergotropy. Since the local subsystems have the same spectrum, the corresponding passive state would be
  \begin{eqnarray}
  \rho^A_{p}=\rho^B_{p} = \lambda_{max} |0\rangle \langle 0| + \lambda_{min} |1\rangle \langle 1|
  \end{eqnarray}
 Using equation (\ref{eq:7}), ergotropic gap of this state turns out to be
  \begin{eqnarray}
   \Delta_{EG} = \lambda_{min}(E_a + E_b)
  \end{eqnarray}
  
 In \cite{Vidal'PRL} $\lambda_{min}$ has been shown to be an entanglement monotone. So $\Delta_{EG}$ should also be a monotone and can be used as a thermodynamic quantifier of entanglement for two-qubit pure states. Maximum value of $\lambda_{min}$ is $\frac{1}{2}$ and the corresponding state is Bell state. This state yields the maximum ergotropic gap among all entangled states and as entanglement decreases ergotropic gap also decreases through $\lambda_{min}$, thereby giving a robust entanglement measure.

  \begin{widetext}
 {\it {\bf Theorem 2:} Consider a $d_1\times d_2$ 
 	bipartite state $\rho^{AB}$ having non-increasing spectrum $\{x_0,x_1,...,x_{d-1}\}$, where $d=d_1d_2$ and without loss of generality, $d_1\leq d_2$, with the marginals governed by linear Hamiltonian. If $\rho^{AB}$  is separable, then ergotropic gap is bounded by 
  
  \begin{eqnarray}
  \Delta_{EG} \leq min\{(Y-Z)E,~M(d_1,d_2)E\},\label{generalcriterion}
  \end{eqnarray} 
   
    where 
    \begin{eqnarray}
    \begin{aligned}
    Y &=\sum\limits_{i=0}^{d_1-1}ix_i + \sum\limits_{i=0}^{d_2-1}ix_i + (d_1-1)\sum\limits_{i=d_1}^{d-1}x_i + (d_2-1)\sum\limits_{i=d_2}^{d-1}x_i \nonumber \\
    Z &=\sum\limits_{i=0}^{d_1-1}i\sum\limits_{k'=0}^{i}x_{ \{\frac{i(i+1)}{2}+k'\}}+\sum\limits_{i=1}^{d_2-d_1}(i+d_1-1)\sum\limits_{k'=1}^{d_1}x_{\{D_1+d_1(i-1)+k'\}}+\sum\limits_{i=1}^{d_1-1}(i+d_2-1)\sum\limits_{j'=1}^{d_1-i}x_{\{D_2+d_1(i-1)-\frac{i(i-1)}{2}+j'\}}. \nonumber
     \end{aligned}
    \end{eqnarray}
  and
  \begin{eqnarray}
  D_1= \frac{d_1(d_1-1)}{2}+(d_1-1)~~~ and ~~~
  D_2=D_1+(d_2-d_1)d_1 \nonumber 
  \end{eqnarray}
  $M(d_1,d_2)$ = \Bigg\{ \begin{tabular}{ccc}
   $\frac{d_1-1}{2}+\frac{d_2-1}{2} - \frac{l}{d_2}[\frac{l^2-1}{3}+m+1]$ ~~~~~if & $(d_2-1) \leq D_1$ \\
   $\frac{d_1+d_2}{2}-1-\frac{d_1}{d_2}[\frac{d^2_1-1}{3}+(k-1)(d_1-1+\frac{k}{2})]- \frac{j(d_1-1+k)(j+1)}{2d_2}$ & if & $d_2-1 > D_1.$
   \end{tabular}\\\\
   
    The interger value of $(l,m)$ and $(k,j)$ will be uniquely determined by the constraint $\frac{l(l+1)}{2}+m = d_2-1$  where $0 \leq m \leq l$ and $D_1+(k-1)d_1+j=d_2-1$; $1\leq j \leq d_1$.}\\
   
  {\it Proof:} Proof has been discussed in the Appendix.
  \end{widetext} 
   {\it {\bf Corollary 2.1:} If $\rho^{AB}$ is a separable two qubit state with the spectrum $(x_0,x_1,x_2,x_3)$ in non-increasing order,  where the reduced subsystems are governed by the same Hamiltonian ${H_{A/B}}=E|1\rangle\langle1|$, then the ergotropy gap is bounded by 
   \begin{equation}
   \Delta_{EG} \leq min \{(x_1 + x_2)E,~~ \frac{E}{2}\}.
   \end{equation}
   Here $\frac{E}{2}$ is the maximum ergotropic gap over the whole state space of separable states.}\\

  {\it Proof:} We will first give an independent proof which has been partitioned as follows.\\
  {\it Spectral dependent criterion}:\\
  Ergotropic gap of a system is given by equation $(\ref{eq:7})$
  \begin{eqnarray}
  \Delta_{EG}= ( p_1 + q_1 )E-( x_1 + x_2 + 2x_3)E
   \label{eq:ergo}
  \end{eqnarray}
 where $\rho^A \equiv (p_0,p_1)$ and $\rho^B \equiv (q_0,q_1)$ are the spectrum of subsystems arranged in non-increasing order.
According to Nielsen-Kempe separable criterion (\ref{NK}),
\begin{eqnarray}
\begin{aligned}
SEP & \implies p_0 \geq x_0 ~,~ q_0 \geq x_0\\
& p_1 \leq (x_1+x_2+x_3) ~,~ q_1 \leq (x_1+x_2+x_3) \label{eq:inequality}
\end{aligned}
\end{eqnarray}
Substituting  inequality (\ref{eq:inequality}) in Eq. (\ref{eq:ergo}) we get
\begin{eqnarray}\label{SEP}
SEP \implies \Delta_{EG}\leq (x_1+x_2)E.
\end{eqnarray}
{\it Dimension dependent criterion}:\\
The above substitution also gives $\Delta_{EG} \leq (p_1 - x_3)E$, which when maximized over all separable states yields the bound
 \begin{eqnarray}
 \begin{aligned}
 \Delta_{EG} & \leq max(p_1 - x_3)E  \nonumber\\
 & = max (p_1E) - min (x_3E)\nonumber\\
  & = \frac{E}{2}.
\end{aligned}
 \end{eqnarray}
 Since $p_0 \geq p_1$ and $x_0 \geq x_1 \geq x_2 \geq x_3$, hence maximum value of $p_1$ is $\frac{1}{2}$ while minimum value of $x_3$ is $0$. So among all separable states, the maximum value of ergotropic gap obtained from dimension dependent criterion is $\frac{E}{2}$.
 \par
 
 Thus, from the above two cases, a necessary criterion for a separable state is that its ergotropic gap should be bounded by
 \begin{equation*}
    \Delta_{EG} \leq min \{(x_1 + x_2)E, \frac{E}{2}\}.
    \end{equation*}

    Alternatively, this result can also be obtained as a special case of Theorem 2 by making the following substitutions:\\
     \begin{eqnarray}
     \begin{aligned}
     d_1&=d_2=2, d=4, \nonumber\\
     ~~D_1&=\frac{d_1(d_1-1)}{2}+(d_1-1)=2 \nonumber\\
      D_2&=D_1+(d_2-d_1)d_1=2
      \end{aligned}
      \end{eqnarray}
      \begin{eqnarray}
     \begin{aligned}
      Y&= \sum\limits_{i=0}^{1}ix_i+\sum\limits_{i=0}^{1}ix_i + \sum\limits_{i=2}^{3}x_i+\sum\limits_{i=2}^{3}x_i=2(x_1+x_2+x_3)\\
      Z&=\sum\limits_{i=0}^{1}i\sum\limits_{k'=0}^{i}x_{ \{\frac{i(i+1)}{2}+k'\}}+\sum\limits_{i=1}^{1}(i+1)\sum\limits_{k'=1}^{2-i}x_{\{2+2(i-1)+k'\}}\\
      & =(x_1+x_2)+2x_3 \label{eq:y-z}
      \end{aligned}
     \end{eqnarray}
     Therefore, $Y-Z=(x_1+x_2)$, which is the spectral dependent criterion.\\
      Now since it follows that $d_2-1 < D_1$, we have 
      \begin{equation}
       M(d_1,d_2)=\frac{d_1+d_2}{2}-1-\frac{l}{d_2}[(l^2-1)+m+1]\label{eq:M(2,2)}.
      \end{equation}
      
 The constraint
 \begin{eqnarray}
 \frac{l(l+1)}{2}+m=1; 0 \leq m \leq l
 \end{eqnarray} would give $(l,m) \equiv (1,0)$ uniquely. Putting these values in (\ref{eq:M(2,2)}) yields $M(2,2)= \frac{1}{2}$.

      \begin{figure}[t!]
         \centering
         	\includegraphics[height=4.5cm,width=5cm]{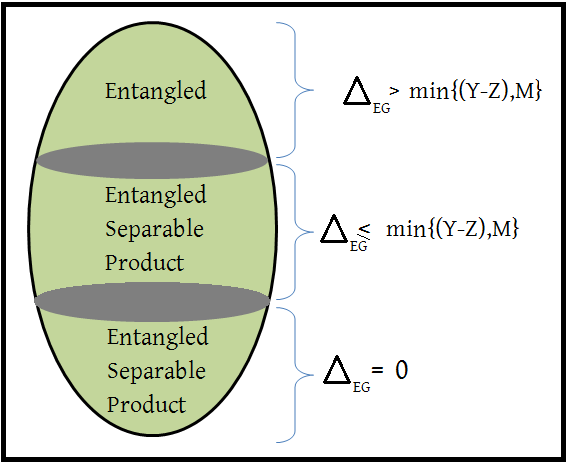}
         	\caption{(Color on-line) The ergotropic gap for all product, separable and some entangled states lie in the interval $[0, \min\{Y-Z,M\}]$. The value beyond this bound gives genuine quantum advantage coming only from entangled states.}\label{fig}
         \end{figure}

{\it {\bf Corollary 2.2:} A two-qubit state with maximally disordered marginals governed by the same Hamiltonian ${H} = E|1\rangle \langle1|$, is separable if and only if 
	\begin{equation}\label{maxdisorder}
\Delta_{EG} \leq (x_1+x_2)E 
	\end{equation}}

{\it Proof:} The proof utilizes the well known result \cite{horo1996PRA} which states that 
 any two-qubit state $\rho$ whose subsystems have maximal entropy is separable iff $x_i \in [0,\frac{1}{2}]$ where $x_i$ is the spectrum of $\rho$ .
\par

Considering the spectrum in non-increasing order, $x_0 \geq x_1 \geq x_2 \geq x_3$,  $SEP \Leftrightarrow x_0 \leq \frac{1}{2}$. If the state is separable Nielsen-Kempe criterion (\ref{NK}) says that, $\frac{1}{2} \geq x_0 \Leftrightarrow 1 \leq 2(x_1+x_2+x_3)$. For marginals with maximally disordered systems ergotropic gap equation (\ref{eq:ergo}) would be, $\Delta_{EG} = E - (x_1+x_2+2x_3)E$. Substituting the above inequality in ergotropic gap we achieve a necessary criterion for separability
\begin{equation}
SEP \implies \Delta_{EG}\leq (x_1+x_2)E.
\end{equation}

To show the sufficiency, take 
\begin{eqnarray}
\begin{aligned}
\Delta_{EG} & \leq (x_1+x_2)E \nonumber \\
1 - (x_1+x_2+2x_3) & \leq (x_1+x_2) \nonumber \\
   (x_0-x_3) & \leq 1-x_0-x_3 \nonumber \\
   x_0 & \leq \frac{1}{2} \Rightarrow SEP
   \end{aligned}
  \end{eqnarray}
\\
Therefore, $\Delta_{EG} \leq (x_1+x_2) \Rightarrow SEP$. Thus for this special class of states, it is a necessary and sufficient criterion  for separability, just like the $\alpha$ Renyi entropy criteria \cite{horo1996PRA} for two-qubit systems. Violation of the criterion implies that the state is entangled. We now present one example from this class.

{\it {\bf Example:} For Bell-diagonal states the thermodynamic criterion is necessary and sufficient. }\\\\
{\it Proof:}

 \begin{eqnarray} 
 \rho^{AB}= x_0 | \phi^+ \rangle \langle \phi^+ | + x_1 | \phi^- \rangle \langle \phi^- | \nonumber \\+ x_2| \psi^+ \rangle \langle \psi^+ | + x_3| \psi^- \rangle \langle \psi^- | \nonumber
 \end{eqnarray}
  where, $|\phi\rangle^+,|\phi\rangle^-,|\psi\rangle^+,|\psi\rangle^-$ are usual bell states and the spectrum has been taken in decreasing order $(x_0 \geq x_1 \geq x_2 \geq x_3)$. Work contribution from marginals is zero because of saturated randomness. Globally one can reach the passive state by some proper entangling unitary
  \begin{eqnarray}
  \rho^{AB}_{p} = x_0|00\rangle\langle00|+x_1|01\rangle\langle01| \nonumber\\
  +x_2|10\rangle\langle10|+x_3|11\rangle\langle11| \nonumber
  \end{eqnarray}
  Using equation (\ref{eq:7})
 \begin{eqnarray}
\Delta_{EG} = E - (x_1+x_2+2x_3)E.
\end{eqnarray}
 The renowned PPT criterion confirms separability for $x_0 \leq \frac{1}{2}$, or in other words, $x_0 \leq \frac{1}{2}$ is sufficient to confirm separability of a state expressed in the above form. \\
 Now if,
 \begin{eqnarray}
 \begin{aligned}
\Delta_{EG} & \leq (x_1+x_2)E \nonumber\\
 1-(x_1 + x_2 + 2x_3) & \leq x_1+x_2 \nonumber\\
 1 & \leq 2(x_1+x_2+x_3) \nonumber \\
 x_0 & \leq \frac{1}{2} \nonumber
 \end{aligned}
  \end{eqnarray}
  And we have already shown the other direction in (\ref{SEP}). Hence
  \begin{eqnarray}
  x_0 \leq \frac{1}{2}\implies SEP \implies \Delta_{EG} \leq (x_1+x_2)E \nonumber
  \end{eqnarray}
  So
\begin{eqnarray}
SEP \Leftrightarrow \Delta_{EG} \leq (x_1+x_2)E \nonumber
\end{eqnarray} 

\par
If a state violates this condition then it is surely entangled and would give quantum advantage.

\section{Bound on ergotropic gap as a thermodynamic criterion of separability}
 Information is a valuable resource in work extraction \cite{bera'18}, i.e., more information about the global system enhances the ability of extracting work. Accordingly, global operations (GO), global unitaries (GU), LOCC, LO and LU provide the following hierarchy 
 
 \begin{equation*}
 W_{GO} \geq W_{GU}\geq W_{LOCC}\geq W_{LO}\geq W_{LU}.
 \end{equation*}
 
 The difference between $W_{GO}$ and $W_{LOCC}$ defined as work deficit,
 was shown to be equal to entanglement distillation 
 for pure states \cite{Oppenheim'02}. In this article, we have considered the difference between $W_{GU}$ and $W_{LU}$ which is defined as ergotropic gap. A bound is provided on this quantity for separable states (Theorem 2). The entangled states which violate this bound show quantum advantage. At this juncture one may ask whether this potentiality can be utilized for entanglement detection?

{\it {\bf Necessary and sufficient criterion of pure entangled states:}} As proved earlier, for pure states,  $\Delta_{EG}\neq 0$, if and only if the state is entangled.  

\begin{figure}[t!]
	\centering
	\includegraphics[scale=0.4]{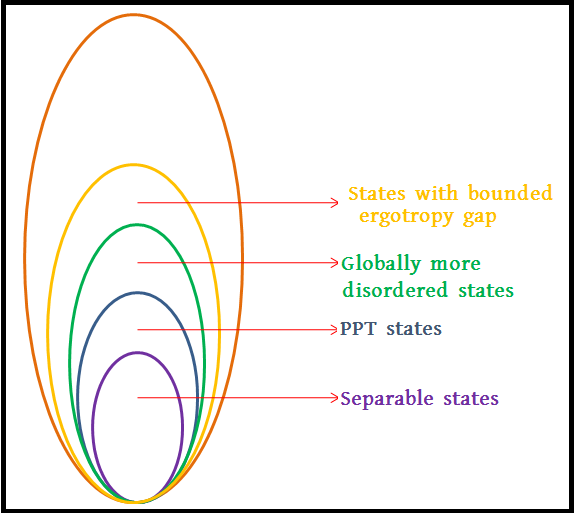}
	\caption{(Color on-line) Outermost convex contour stands for total state space, whereas, the inner most depicts separable region. Others stands for several criteria depending upon their faithfulness to detecting entanglement.}\label{fig}
\end{figure} 

\par Comparison of entanglement between two pure states depends on the task to be performed. For example, $\phi \rightarrow \psi$ transformation is possible under LOCC if and only if  $\lambda(|\phi\rangle)  \prec \lambda(|\psi\rangle) $\cite{Nielsen'99}. So $\phi$ is more entangled than $\psi$ and this implies according to Theorem 1 that ergotropic gap of $|\phi\rangle$ is greater than that of $|\psi\rangle$. However, the converse is not true in general but for two-qubit states we have shown in corollary $1.1$ that as ergotropic gap decreases entanglement also decreases. 
\par
{\it {\bf Sufficient criterion of bipartite mixed entangled states:}}
 According to Theorem 2 if for a bipartite state $\Delta_{EG} > min\{Y-Z,M(d_1,d_2)\}$ then the state is an entangled state. The reduced subsystems are governed by the linear Hamiltonian, and for further discussion we take $E=1$.
 \par
 In corollary 2.1 we found that if $\Delta_{EG} > min\{x_1+x_2, \frac{1}{2}\}$ then the state is surely an entangled state, whereas in corollary 2.2 it was shown that for states having maximally mixed marginals, $\Delta_{EG} > (x_1+x_2)$ becomes necessary and sufficient to characterize entanglement.  
 
 \par
Although { \it thermodynamic criterion} follows from the {\it Nielsen-Kempe disorder criterion (\ref{NK})} we have shown that it becomes necessary and sufficient to exhibit entanglement for Werner class which is a convex mixture of the singlet state with probability $p$ and completely random state with probability $(1-p)$ having spectrum $\rho_{w} \equiv (\frac{1+3p}{4},\frac{1-p}{4},\frac{1-p}{4},\frac{1-p}{4})$. By optimal entangling unitaries $\{|\psi^{-}\rangle \rightarrow |00\rangle, |\psi^{+}\rangle \rightarrow |01\rangle,|\phi^{+}\rangle \rightarrow |10\rangle,|\phi^{-}\rangle \rightarrow |11\rangle\}$ one can achieve $\Delta_{EG} =p$. It is well known that for $p \leq \frac{1}{3}$, this class is separable and our  thermodynamic criterion  given in Eq.(\ref{maxdisorder}) is also satisfied for this entire range. The Bell CHSH inequality \cite{bell'69} used to experimentally detect entanglement, confirms entanglement only for $p > 0.7056$ \cite{Vertesi'08}, whereas our criterion can be implemented experimentally and captures the complete range. On the other hand,  negativity of Von-Neumann conditional entropy $(S(B|A)=S(AB)-S(A))$ of a state is also useful in entanglement detection \cite{horodecki'94}. For Werner class of states it is negative for $p\geq \frac{3}{4}$, whereas our criterion captures more entangled states than this one also. But since our thermodynamic criterion is spectral dependent,  it cannot detect any PPT entangled state. Never the less,  it has its own beauty since it is an operational criterion where we can detect entanglement of a system through thermodynamic work. 
         
   \section{Global operation vs local operation as a criterion of entanglement}    
   In this scenario, Alice and Bob have access to individual local thermal bath  of inverse temperature $\beta$, governed by the Hamiltonian $H^{b}_A/H^{b}_B$ . Now if a state $\rho^{AB}$ is shared between them where Alice and Bob have $\rho^A$ and $\rho^B$ part governed by the Hamiltonian $H_A$ and $H_B$ respectively, they would perform local thermal operations to extract work. The amount of the extracted work will be equal to the corresponding free energy difference between initial($\rho^A/\rho^B$) and final($\tau^A_{\beta}/\tau^B_{\beta}$) state \cite{popescuNAT},
   \begin{equation*}
   W_{LO}= F(\rho^A)-F(\tau^A_{\beta})+ F(\rho^B)-F(\tau^B_{\beta}).
   \end{equation*}
   
   However, when we bring the systems and their corresponding baths together, the joint thermal state $\tau^A_{\beta}\otimes\tau^B_{\beta}$ would be at the same temperature $\beta$. It is assumed that the bath is governed by the interaction free Hamiltonian $H_g= H_A^{b}\otimes I_B + I_A \otimes H_B^{b}$ and corresponding extractable work under global thermal operation would be
    \begin{equation*}
    W_{GO}=F(\rho^{AB})-F(\tau^A_{\beta}\otimes\tau^B_{\beta}).
    \end{equation*}  
    The work difference between global and local operation is defined by
    \begin{equation*}
    \begin{aligned}
     \Delta = W_{GO}-W_{LO} & = F(\rho^{AB})-F(\rho^A)-F(\rho^B) \nonumber \\
                            & = E(\rho^{AB})-\frac{1}{\beta}S(AB)-E(\rho^A)-E(\rho^B)\nonumber \\
                            & +\frac{1}{\beta}\{S(A)+S(B)\} \nonumber \\
                            & =\frac{1}{\beta}\{S(A)+S(B)-S(AB)\} \nonumber \\
                            & = \frac{1}{\beta}I(A:B). 
      \end{aligned}
    \end{equation*}
    The internal energy and entropy of a system is defined by $E(\rho^X)=Tr(\rho^X H_X)$ and $S(\rho^X)=S(X)=-Tr(\rho^Xlog\rho^X)$. If the state $\rho^{AB}$ is separable,
    \begin{equation*}
    \begin{aligned}
   max I(A:B)&= max\{S(A)-S(A|B)\} \nonumber\\ &\leq max_{\rho^A} S(A) - min_{\rho^{AB}} S(A|B) \nonumber \\
              & = logd_A	
    \end{aligned}
    \end{equation*}  
    In the same way it can be shown that $max I(A:B) \leq logd_B$. So if a state is separable, then
    \begin{equation}
    \Delta = \frac{1}{\beta}I(A:B) \leq min\frac{1}{\beta} [logd_A,logd_B].
    \end{equation}
     Mutual information above this bound certify entanglement. For Werner class, this criterion is able to detect entanglement for $p \geq \frac{3}{4}$.  However, it is weaker than negative conditional entropy criterion as well as our ergotropic gap criterion.
\section{Experimental indication} 

We have seen that $\Delta_{EG} > min(x_1+x_2,\frac{1}{2})$ detects entanglement, where $(x_1 + x_2)$ is the spectral dependent criterion and $\frac{1}{2}$ which is the maximum value over all separable two-qubit states, is the dimension dependent criterion. 
Let's understand through some examples, why the above minimization is needed. For  the spectrum  $(\frac{3}{4},\frac{1}{4},0,0)$, the spectral condition gives $\Delta_{EG} > (x_1+x_2) = \frac{1}{4}$, which is sufficient to confirm entanglement. On the other hand, for the spectrum
 $(\frac{1}{3},\frac{1}{3},\frac{1}{3},0)$, the spectral condition yields $(x_1+x_2) = \frac{2}{3}$, but we know through dimension dependent criterion that the maximum bound on $\Delta_{EG}$ over all separable states is $\frac{1}{2}$. That is why we have to take the minimum of the two. But if the state is unknown then  $\Delta_{EG}>\frac{1}{2}$ confirms entanglement and we cannot improve this bound further.

Detection of entanglement can be done for bipartite states $\rho^{AB}$ whose marginals $\rho_p^A$ and $\rho^B_p$ are passive, or in more realistic situation, completely passive (thermal of virtual temperature $\beta_1$ and $\beta_2$ respectively).
So the local ergotropy for these states is zero and the ergotropic gap would be equal to just the global ergotropy.
In order to implement this, we need a continuously varying unitary on the global system. This induces a change in the system's energy and correspondingly there is either work loss or work gain. The maximum work gain is defined as ergotropic gap for passive marginal states. If a system violates the given bound on ergotropic gap, we are sure that the state is entangled, otherwise we can't make any comment.
\par 
For the Werner class, however, only a single apparatus (unitary) is needed to detect the entanglement completely. Since $\Delta_{EG}=p$, one can certify the given Werner state through this value.
\par
One limitation of the experimental set up described above is that for non passive marginals, entanglement detection of an unknown state is not possible . Since ergotropic gap is the difference between global and local ergotropy defined for the optimal unitary only, it can sometimes happen that for a separable state one can cross the optimal bound (\ref{generalcriterion}) by some inappropriate unitary. So experimentally this kind of state can not be detected correctly through the thermodynamic criterion. 

 \begin{figure}[t!]
          	\centering
          	\includegraphics[scale=0.4]{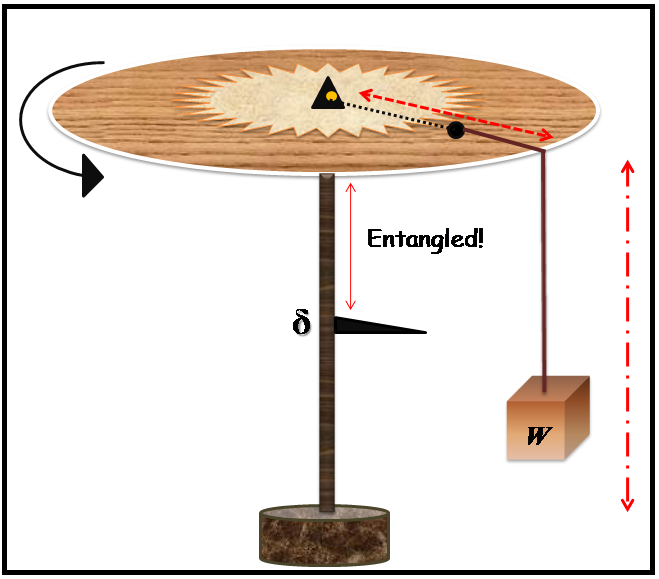}
          \caption{(Color on-line)\textbf{A schematic experimental set-up:} A source of the bipartite quantum state to be tested is placed at the center of the disc. Emitted system particles are captured in the black ball on the disc. Rotation of the disc along it's axis represents the continuous parametric classes of applied unitary.  Work gain or loss of the system is represented by the radially inward or outward motion of the ball respectively, which causes a vertical shift of the work load $W$. Whenever the upward shift crosses the index line $\delta$,  the system state is detected as entangled.}\label{fig}
          \end{figure} 
\section{ Ergotropic Gap as dimension witness}
 We have seen that for any $d\times d$  system, the maximum ergotropic gap comes from a maximally entangled state $|\psi\rangle^{AB}=\sum\limits_{i=0}^{d-1}\frac{1}{\sqrt{d}}|ii\rangle$.
From equation (\ref{eq:7}), 
\begin{eqnarray}
\Delta_{EG}=Tr(\rho^A_p{H_A})+Tr(\rho_p^B{H_B})=(d-1)
\end{eqnarray}

    If the ergotropy gap of a given state is greater than $d-1$, then the system dimension would be at least $d+1$. Thus, ergotropic gap gives a lower bound on the dimension of the system and hence acts as a dimension witness. For example,  if $\Delta_{EG}$ is 1.5, then the system dimension would be at least $3\times 3$.
    
   \begin{figure}[!htb]
            \centering
            	\includegraphics[scale=0.4]{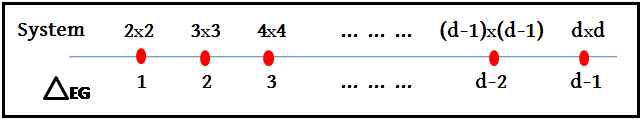}
            	\caption{(Color on-line) Red dots are the bound on ergotropic gap for correponding systems and defined by positive integers.  }\label{fig}
            \end{figure} 
\section{Conclusion}

We have proposed an operational task which can detect a large class of bipartite entangled states depending upon the difference between globally and locally extractable work by unitary operations.  For an arbitrary pure multipartite quantum system, non-zero ergotropic gap is a necessary and sufficient condition to guarantee entanglement. It has also been established that the majorization criterion of state transformation for bipartite pure entangled states has a direct connection to the hierarchy of ergotropic gap. For any arbitrary bipartite separable state our task provides a bound, beyond which the state can be certified as entangled. The criterion is derived as a consequence of an operational task and the experimental realization is valid for the class of states whose marginals are locally thermal at some given temperature. This gives a physical interpretation of the well known Nielsen-Kempe disorder criterion. We have also shown that the difference in extractable work by GO and LO is bounded by the quantum mutual information between the subsystems.  
Another interesting point is that the bound on ergotropic gap  provides a dimension witness for all $d\times d$ quantum states.

As a generalization of our work, it would be interesting to obtain the thermodynamic bound on the ergotropic gap in the  multi-partite scenario. Although there exist some statistical criteria, which invoke measurement cost to detect bi-separability and genuineness of entanglement in multiparty scenario, bounds on ergotropic gap may provide sharper classification. For two-qubit states our criterion captures the same region of state space as Tsalli's and Renyi entropy criterion and it is a subject of further research to compare this for a general bipartite state.

\section*{Acknowledgement}
M.A. acknowledges financial support from the CSIR project 09/093(0170)/2016-EMR-I.

\begin{widetext}

\section*{Appendix}

  \section*{ Proof of Theorem 2} Consider a $d_1\times d_2$ quantum state $\rho^{AB},$ with spectrum $(x_0,x_1,\cdots,x_{d-1})$ having marginals as  $\rho^A = Tr_B(\rho^{AB}) \equiv (p_0,p_1,\cdots,p_{d_1-1})$ and $\rho^B = Tr_A(\rho^{AB}) \equiv (q_0,q_1,\cdots,q_{d_2-1})$. The spectrum is arranged in non-increasing order and $d=d_1d_2$, where we assume that $d_1 \leq d_2$. The proof involves two parts: the spectral dependent and the dimension dependent criterion.
  
    \subsection {Spectral dependent criterion}
    	
    Ergotropy gap of the  system is given by equation (\ref{eq:7}),
    \begin{eqnarray}\label{ap1}
    \Delta_{EG} = \sum\limits_{i=1}^{d_1-1}ip_{i} + \sum\limits_{i=1}^{d_2-1}iq_{i} - Tr(\rho^{AB}_{p} H_{AB})
    \end{eqnarray}
  
    where the summations are the passive state energy of $A$ and $B$ respectively and the last term is the global passive state energy.
    
    If the state $\rho^{AB}$ is separable then the following condition (Nielse-Kempe disorder criterion) holds:
    \begin{eqnarray}
    \lambda(\rho^A)\succ \lambda(\rho^{AB})~~~ \wedge~~~ \lambda(\rho^B)\succ \lambda(\rho^{AB}), \nonumber
    \end{eqnarray}
where $\lambda$ denotes the spectrum of the corresponding states.
    
    \begin{align}\label{ap2}
      \left.
      \begin{array}{r@{}l}
        p_0 \geq x_0\\
        p_0 + p_1 \geq x_0 + x_1\\
        .\\
        .\\
        .\\
       p_0+...+p_i \geq x_0+...+x_i \\
       .\\.\\.\\
       \sum\limits_{i=0}^{d_1-2} p_i \geq \sum\limits_{i=0}^{d_1-2} x_i
        & 
      \end{array}
      \right\rbrace  \left\lbrace
      \begin{array}{r@{}l}
      q_0 \geq x_0\\
          q_0 + q_1 \geq x_0 + x_1\\
          .\\
          .\\
          .\\
         q_0+...+q_i \geq x_0+...+x_i\\
            .\\.\\.\\
            \sum\limits_{i=0}^{d_2-2} q_i \geq \sum\limits_{i=0}^{d_2-2} x_i 
         &
      \end{array}
      \right.
    \end{align}
     can be expressed as\\
     \begin{align}\label{ap3}
       \left.
       \begin{array}{r@{}l}
         \sum\limits_{i=1}^{d_1-1}p_i \leq \sum\limits_{i=1}^{d-1}x_i\\
         \sum\limits_{i=2}^{d_1-1}p_i \leq \sum\limits_{i=2}^{d-1}x_i\\
         .\\
         .\\
         .\\
        p_{i+1}+...+p_{d_1-1} \leq x_{i+1}+...+x_{d-1} \\
        .\\.\\.\\
        p_{d_1-1} \leq \sum\limits_{i=d_1-1}^{d-1}x_i
         & 
       \end{array}
       \right\rbrace  \left\lbrace
       \begin{array}{r@{}l}
        \sum\limits_{i=1}^{d_2-1}q_i \leq \sum\limits_{i=1}^{d-1}x_i\\
            \sum\limits_{i=2}^{d_2-1}q_i \leq \sum\limits_{i=2}^{d-1}x_i\\
            .\\
            .\\
            .\\
           q_{i+1}+...+q_{d_2-1} \leq x_{i+1}+...+x_{d-1} \\
           .\\.\\.\\
           q_{d_2-1} \leq \sum\limits_{i=d_2-1}^{d-1}x_i
          &
       \end{array}
       \right.
     \end{align}
     
     Adding the above inequalities gives
     \begin{eqnarray} \label{ap4}
      \sum\limits_{i=1}^{d_1-1}ip_{i} + \sum\limits_{i=1}^{d_2-1}iq_{i} \leq \sum\limits_{i=1}^{d_1-1}ix_{i} + \sum\limits_{i=1}^{d_2-1}ix_{i}+  (d_1-1)\sum\limits_{i=d_1}^{d-1}x_{i}+(d_2-1)\sum\limits_{i=d_2}^{d-1}x_{i},
     \end{eqnarray}
    
    which when substituted in Eq.(\ref{ap1}) yields
    \begin{eqnarray}
    \Delta_{EG} \leq \sum\limits_{i=1}^{d_1-1}ix_{i} + \sum\limits_{i=1}^{d_2-1}ix_{i}+  (d_1-1)\sum\limits_{i=d_1}^{d-1}x_{i}+(d_2-1)\sum\limits_{i=d_2}^{d-1}x_{i} - Tr(\rho^{AB}_{p} H_{AB}) \nonumber \\
    = Y - Z.\label{ap5}
    \end{eqnarray}
    
    The local passive state energy is bounded by 
    \begin{eqnarray}
    Y = \sum\limits_{i=1}^{d_1-1}ix_{i} + \sum\limits_{i=1}^{d_2-1}ix_{i}+  (d_1-1)\sum\limits_{i=d_1}^{d-1}x_{i}+(d_2-1)\sum\limits_{i=d_2}^{d-1}x_{i} \label{ap6}
    \end{eqnarray}  and
    \begin{eqnarray}
    Z = Tr(\rho^{AB}_{p} H_{AB})
    \end{eqnarray}is the corresponding global passive state energy.

    To make the calculation of $Z$ easier,  we consider a matrix $X$ by taking the whole global spectrum, where off diagonal elements have the same energy (denoted by same color),
    viz., $|01\rangle$ and $|10\rangle$ states have same energy. We designate their corresponding spectrum in passive state as
    \begin{eqnarray}
    x_1 \rightarrow |01\rangle ~~;~~ x_2 \rightarrow |10\rangle \nonumber
    \end{eqnarray}
      In the same way, 
      \begin{eqnarray}
      x_3 \rightarrow |02\rangle ~~;~~ x_4 \rightarrow |11\rangle ~~;~~ x_5 \rightarrow |20\rangle \nonumber
      \end{eqnarray}
      The above successive indices have equal energy and are situated on the next off diagonal of $X$ (indicated in blue).
        
\begin{equation}
  \mathbf{X}=
  \begin{blockarray}{*{9}{c} l}
    \begin{block}{*{9}{>{$\footnotesize}c<{$}} l}
    \textcolor{green}{0}&\textcolor{red}{1}&~\textcolor{blue}{2} &$\cdots$ &  & \textcolor{cyan}{$l'$}
      & &$\cdots $  & \textcolor{magenta}{$d_{1}-1$}
       \\
    \end{block}
    \begin{block}{[*{9}{c}]>{$\footnotesize}l<{$}}
     \textcolor{green}{x_0} & \textcolor{red}{x_1} &~~~ \textcolor{blue}{x_3} &  & & \textcolor{cyan}{x_{\frac{l'(l'+1)}{2}}} & & & \textcolor{magenta}{x_{\frac{d_1(d_1-1)}{2}}} &  \\
     \textcolor{red}{x_2} & \textcolor{blue}{x_4} &&&&&&& \textcolor{brown}{x_{D_1+1}} & $~\textcolor{brown}{d_1-1+1}$\\
     \textcolor{blue}{x_5} &&&&&&&&$\vdots$ &  \textcolor{gray}{$d_1-1+2$}\\ 
      $\vdots$ &&\textcolor{cyan}{x_{\frac{l'(l'+1)}{2}+m'}}&&&&&& $\vdots$ &  \\
       $\vdots$ &&&&&&&& \textcolor{orange}{x_{D_1+(k'-1)d_1+1}} & \textcolor{orange}{$d_1-1+k'$}  \\
        $\vdots$ &&&&&&&& $\vdots$ & \\
     $\vdots$ &&&&&&&&\textcolor{teal}{x_{D_1+(d_2-d_1)d_1-d_1+1}}& $\textcolor{teal}{d_2-1}$  \\
     \textcolor{cyan}{x_{\frac{l'(l'+1)}{2}+l'}} &&&&&&&& $\vdots$ & \\
     $\vdots$ &&& \textcolor{orange}{x_{D_1+(k'-1)d_1+j'}}&&&&& $\vdots$ &  \\
     $\vdots$ &&&&&&&& $\vdots$ & \\
     \textcolor{magenta}{x_{D_1=\frac{d_1(d_1-1)}{2}+d_1-1}}& \textcolor{brown}{x_{D_1+(d_1-1)}} &&&&&& & \textcolor{black}{x_{d^2_1 -1}}& ~$\textcolor{black}{2d_1-2}$\\
     \textcolor{brown}{x_{D_1+d_1}} &&&&&&& & $\vdots$&\\
      $\vdots$ &&&&&&&& \textcolor{violet}{x_{D_2+(i-1)d_1-\frac{i(i-1)}{2}+1}} & ~$\textcolor{violet}{d_2-1+i}$ \\
      \textcolor{orange}{x_{D_1+k'd_1}} &&&&&&&& $\vdots$ &  \\
         $\vdots$&   &&&&&& &&\\
           \textcolor{teal}{x_{D_2=D_1+(d_2-d_1)d_1}}& &\textcolor{violet}{x_{D_2+(i-1)d_1-\frac{i(i-1)}{2}+(d_1-i)}}&&&&& &\textcolor{olive}{x_{d-1}}&~$\textcolor{olive}{d_1+d_2-2}$\\
     \end{block}
  \end{blockarray}
  \end{equation}
  
  In general,the off-diagonal $\{ \textcolor{cyan}{x_{\frac{l'(l'+1)}{2}}}  \textcolor{cyan}{x_{\frac{l'(l'+1)}{2}+1}},  \cdots ,\textcolor{cyan}{x_{\frac{l'(l'+1)}{2}+m'}}, \cdots ,\textcolor{cyan}{x_{\frac{l'(l'+1)}{2}+l'}} \}$ has energy $\textcolor{cyan}{l'}$ with the constraint $ 0\leq \textcolor{cyan}{m'}\leq \textcolor{cyan}{l'}$.
  \par
     Passive state energy for any arbitrary state $\rho^{AB}$ is given by
     \begin{eqnarray}{\label{ap7}}
     \begin{aligned}
    Z = Tr(\rho^{AB}_{p}H_{AB}) & = \sum\limits_{i=1}^{d_1-1}i\sum\limits_{k'=0}^{i}x_{\frac{i(i+1)}{2}+k'} + \sum\limits_{k'=1}^{d_2-d_1}(d_1-1+k')\sum\limits_{j'=1}^{d_1}x_{D_1+(k'-1)d_1+j'} \\  
    &+ \sum\limits_{i=1}^{d_1-1}(d_2-1+i)\sum\limits_{j'=1}^{d_1-i}x_{D_2+(i-1)d_1-\frac{i(i-1)}{2}+j'} 
          \end{aligned}
     \end{eqnarray}
      where, the $1^{st}$ term on the right side of equality (\ref{ap7}) gives us the energy due for the spectrum $\{x_0,x_1,x_3, \cdots \cdots , x_{D_1}\}$, $2^{nd}$ term calculate the contribution for the spectrum $\{x_{D_1+1},x_{D_1+2}, \cdots \cdots , x_{D_2}\}$ and the rest of the spectrum $\{x_{D_2+1},x_{D_2+2}, \cdots,x_{d-1}\}$ contributes in the last term.
      
      Substituting the above expression for $Z$ in Eq.(\ref{ap5}) gives the bound on ergotropic gap for separable states
        \begin{eqnarray}
        \begin{aligned}
         \Delta_{EG} & \leq  \sum\limits_{i=1}^{d_1-1}ix_{i} + \sum\limits_{i=1}^{d_2-1}ix_{i}+  (d_1-1)\sum\limits_{i=d_1}^{d-1}x_{i}+(d_2-1)\sum\limits_{i=d_2}^{d-1}x_{i} \\&-\sum\limits_{i=1}^{d_1-1}i\sum\limits_{k'=0}^{i}x_{\frac{i(i+1)}{2}+k'}-\sum\limits_{k'=1}^{d_2-d_1}(d_1-1+k')\sum\limits_{j'=1}^{d_1}x_{D_1+(k'-1)d_1+j'}
         -\sum\limits_{i=1}^{d_1-1}(d_2-1+i)\sum\limits_{j'=1}^{d_1-i}x_{D_2+(i-1)d_1-\frac{i(i-1)}{2}+j'}
         \end{aligned}
          \end{eqnarray}
     
      \subsection{ Dimension dependent criterion}
      
       Dimension dependent criterion gives the bound on ergotropic gap for all separable states in a given arbitrary dimension. Let us proceed with the proof.
        
        {\bf CASE-I : $d_2-1 \leq D_1 = \frac{d_1(d_1-1)}{2}+(d_1-1)$}\\
        Rewriting equation (\ref{ap1})
        \begin{eqnarray}\label{ap9}
        \begin{aligned}
        \Delta_{EG} & =\sum\limits_{i=1}^{d_1-1}ip_{i} + \sum\limits_{i=1}^{d_2-1}iq_{i} -\sum\limits_{i=1}^{l-1}i\sum\limits_{k'=0}^{i}q_{\{\frac{i(i+1)}{2}+k'\}}-l\sum\limits_{k'=0}^{m}q_{\{\frac{l(l+1)}{2}+k'\}} \\ & +\sum\limits_{i=1}^{l-1}i\sum\limits_{k'=0}^{i}q_{\{\frac{i(i+1)}{2}+k'\}}+l\sum\limits_{k'=0}^{m}q_{\{\frac{l(l+1)}{2}+k'\}}- Tr(\rho^{AB}_{p} H_{AB})
        \end{aligned}
        \end{eqnarray}
        
       where a term has been added and subtracted on purpose.

      Now we choose $l$ number of inequalities of $q's$ from Eq.(\ref{ap3})

              \begin{eqnarray}\label{ap10}
              \begin{aligned}
              1.~~~ q_{1}+...+q_{d_2-1} &\leq x_{1}+...+x_{d-1}  \\
              2.~~~ q_{3}+...+q_{d_2-1} &\leq x_{3}+...+x_{d-1} \\
              3.~~~ q_{6}+...+q_{d_2-1} &\leq x_{6}+...+x_{d-1}  \\
              \vdots  \\
               l-1.~~~ q_{\frac{l(l-1)}{2}}+...+q_{d_2-1} &\leq x_{\frac{l(l-1)}{2}}+...+x_{d-1} \\
               l.~~~ q_{\frac{l(l+1)}{2}}+...+q_{d_2-1} &\leq x_{\frac{l(l+1)}{2}}+...+x_{d-1}
                \end{aligned}
              \end{eqnarray}
      
      We can visualize these by putting them in matrix form as follows
        
        \begin{equation}\label{ap11}
          \mathbf{Q1}=
          \begin{blockarray}{*{9}{c} l}
            \begin{block}{*{9}{>{$\footnotesize}c<{$}} l}
            \textcolor{green}{0}&\textcolor{red}{1}&~~~\textcolor{blue}{2} & $\cdots$ & $\cdots$ & \textcolor{cyan}{l}
              &$\cdots$ &$\cdots $  & \textcolor{magenta}{$d_1-1$}
               \\
            \end{block}
            \begin{block}{[*{9}{c}]>{$\footnotesize}l<{$}}
             \textcolor{green}{q_0} & \textcolor{red}{q_1} &~~~ \textcolor{blue}{q_3} &  & & \textcolor{cyan}{q_{\frac{l(l+1)}{2}}} & & & 0 &  \\
             \textcolor{red}{q_2} & \textcolor{blue}{q_4} &&&&&&& 0 &\\
             \textcolor{blue}{q_5} &&&&&&&&$\vdots$ &\\ 
              $\vdots$ &&\textcolor{cyan}{q_{\frac{l(l+1)}{2}+m}}=\textcolor{cyan}{q_{d_2-1}}&&&&&& $\vdots$ &  \\
             $\vdots$ &0&&&&&&& $\vdots$ &  \\
             0 &&&&&&&& $\vdots$ & \\
             $\vdots$ &&&&&&&& $\vdots$ &  \\
             $\vdots$ &&&&&&&& $\vdots$ & \\
             0 &&&&&&& &0&\\
             \end{block}
          \end{blockarray}
          \end{equation}
      Let
      \begin{eqnarray}\label{ap12}
      R=1.(q_1+q_2)+2.(q_3+q_4+q_5)+ \cdots + (l-1)[q_{\frac{l(l-1)}{2}}+\cdots + q_{\frac{l(l-1)}{2}+(l-1)}]+l[q_{\frac{l(l+1)}{2}}+ \cdots + q_{\frac{l(l+1)}{2}+m}].
      \end{eqnarray}
     Since $(l,m)$ are integers, they are determined uniquely from the condition
     \begin{eqnarray}\label{ap13}
     \frac{l(l+1)}{2}+m=d_2-1~~~~ where ~~~~0 \leq m \leq l.
     \end{eqnarray}
     
     If we take sum of all the inequalities (\ref{ap10}), then the L.H.S would eventually give back equation (\ref{ap12}),
     \begin{eqnarray}\label{ap14}
    R = \sum\limits_{i=1}^{l-1}i\sum\limits_{k'=0}^{i}q_{\{\frac{i(i+1)}{2}+k'\}}+l\sum\limits_{k'=0}^{m}q_{\{\frac{l(l+1)}{2}+k'\}}.
     \end{eqnarray}
     
     and the R.H.S is
       \begin{eqnarray}\label{ap15}
       \begin{aligned}
       R'&=1.(x_1+x_2)+2.(x_3+x_4+x_5)+ \cdots + (l-1)[x_{\frac{l(l-1)}{2}}+\cdots + x_{\frac{l(l-1)}{2}+(l-1)}]+l[x_{\frac{l(l+1)}{2}}+ \cdots + x_{d-1}] \\
       &=\sum\limits_{i=1}^{l-1}i\sum\limits_{k'=0}^{i}x_{\frac{i(i+1)}{2}+k'}+l\sum\limits_{k'=\frac{l(l+1)}{2}}^{d-1}x_{k'}
        \end{aligned}
       \end{eqnarray}
      Since $R \leq R'$, substituting the values of $+R$ by $+R'$ and expression for $Z$ (Eq.(\ref{ap7}))  in Eq.(\ref{ap9}) we get 
       
       \begin{eqnarray}\label{ap16}
       \begin{aligned}
       \Delta_{EG} & \leq \sum\limits_{i=1}^{d_1-1}ip_{i} + \sum\limits_{i=1}^{d_2-1}iq_{i} -\sum\limits_{i=1}^{l-1}i\sum\limits_{k'=0}^{i}q_{\{\frac{i(i+1)}{2}+k'\}}-l\sum\limits_{k'=0}^{m}q_{\{\frac{l(l+1)}{2}+k'\}} +\sum\limits_{i=1}^{l-1}i\sum\limits_{k'=0}^{i}x_{\frac{i(i+1)}{2}+k'}+l\sum\limits_{k'=\frac{l(l+1)}{2}}^{d-1}x_{k'} \\ &-[\sum\limits_{i=1}^{d_1-1}i\sum\limits_{k'=0}^{i}x_{\frac{i(i+1)}{2}+k'} + \sum\limits_{k'=1}^{d_2-d_1}(d_1-1+k')\sum\limits_{j'=1}^{d_1}x_{D_1+(k'-1)d_1+j'}+ \sum\limits_{i=1}^{d_1-1}(d_2-1+i)\sum\limits_{j'=1}^{d_1-i}x_{D_2+(i-1)d_1-\frac{i(i-1)}{2}+j'}]
        \end{aligned}
       \end{eqnarray}
              
              Now since 
              \begin{eqnarray}
              \frac{l(l+1)}{2}+m = d_2-1 \leq \frac{d_1(d_1-1)}{2}+(d_1-1) ~~~;~~~ l < d_1 \nonumber
              \end{eqnarray}
               so the term in the parentheses  is always greater than the sum of $5^{th}$ and $6^{th}$ term. Hence
               
                \begin{equation}\label{ap17}
                \Delta_{EG} \leq \sum\limits_{i=1}^{d_1-1}ip_{i} + \sum\limits_{i=1}^{d_2-1}iq_{i} -\sum\limits_{i=1}^{l-1}i\sum\limits_{k'=0}^{i}q_{\{\frac{i(i+1)}{2}+k'\}}-l\sum\limits_{k'=0}^{m}q_{\{\frac{l(l+1)}{2}+k'\}}-\delta.
                \end{equation}
                Distributing the $2^{nd}$ term we obtain
                
           \begin{equation}\label{ap18}
           \Delta_{EG} \leq \sum\limits_{i=1}^{d_1-1}ip_{i} +\sum\limits_{i=1}^{\frac{l(l+1)}{2}-1}(i-l')q_i+\sum\limits_{i=\frac{l(l+1)}{2}}^{d_2-1}(i-l)q_i-\delta
           \end{equation}
           where $(l',m')$ are fixed by the value of $i$, following the constraint $i=\frac{l'(l'+1)}{2}+m'$; $0 \leq m' \leq l'$. In this expression, frequency of $q_i$ is $(i-l')$ and $(i-l)$ which increase as $i$ increases.
           \par
           
           Now we find the spectrum which would give the maximum value of the bound on ergotropic gap for separable states. So maximizing Eq.(\ref{ap18}) we obtain
           
           \begin{eqnarray}\label{ap19}
           \begin{aligned}
           \Delta_{EG} \leq M(d_1,d_2) & = max (\Delta_{EG})\\
            & = max_{p_i} (\sum\limits_{i=1}^{d_1-1}ip_{i}) + max_{q_i}(\sum\limits_{i=1}^{\frac{l(l+1)}{2}-1}(i-l')q_i) +  max_{q_i}(\sum\limits_{i=\frac{l(l+1)}{2}}^{d_2-1}(i-l)q_i)- min (\delta)
           \end{aligned}
           \end{eqnarray}
           We know that $min(\delta)=0.$
           Since the frequency increases as $i$ increases, we would choose maximum value $p_i$ and $q_i$ for greater $i$. As $p_i's$ and $q_i's$ are defined in non increasing order, the best choice is the uniform distribution
           \begin{eqnarray}\label{ap20}
           (p_1,p_2, \cdots, p_{d_1-1})\equiv (\frac{1}{d_1},\frac{1}{d_1}, \cdots,\frac{1}{d_1})\nonumber \\
                      (q_1,q_2, \cdots, q_{d_2-1})\equiv (\frac{1}{d_2},\frac{1}{d_2}, \cdots,\frac{1}{d_2}). \nonumber \\
              \end{eqnarray}
              So maximally disordered marginals give the maximum ergotropic gap over all separable states. Substituting (\ref{ap20}) in Eq.(\ref{ap17}) or (\ref{ap19}) one can easily get
          
          \begin{equation*}
          \Delta_{EG} \leq \frac{d_1-1}{2}+\frac{d_2-1}{2}-\frac{1}{d_2}[\frac{l(l-1)}{2}+\frac{l(l-1)(2l-1)}{6}]-\frac{1}{d_2}[l(m+1)] = M(d_1,d_2)
          \end{equation*}
          \begin{equation}\label{ap21}
          M(d_1,d_2) = \frac{d_1-1}{2}+\frac{d_2-1}{2} - \frac{l}{d_2}[\frac{l^2-1}{3}+m+1]
          \end{equation}
          
           {\bf Case II: $d_2-1 > D_1 = \frac{d_1(d_1-1)}{2}+(d_1-1)$}\\
           
           Following the separability criterion $(d_1-1+k)$ number of $q_i's$ inequalities have been chosen from inequality (\ref{ap3})
            \begin{eqnarray} \label{ap22}
            \begin{aligned}
             1.~~~ q_{1}+...+q_{d_2-1} &\leq x_{1}+...+x_{d-1} \\
             2.~~~ q_{3}+...+q_{d_2-1} &\leq x_{3}+...+x_{d-1} \\
             3.~~~ q_{6}+...+q_{d_2-1} &\leq x_{6}+...+x_{d-1} \\
             \vdots & \\
              d_1-1.~~~ q_{\frac{d_1(d_1-1)}{2}}+...+q_{d_2-1} &\leq x_{\frac{d_1(d_1-1)}{2}}+...+x_{d-1}\\
               d_1.~~~ q_{\frac{d_1(d_1+1)}{2}}+...+q_{d_2-1} &\leq x_{\frac{d_1(d_1+1)}{2}}+...+x_{d-1}\\
               (d_1+1).~~~ q_{D_1+d_1+1}+...+q_{d_2-1} &\leq x_{D_1+d_1+1}+...+x_{d-1} \\
               (d_1+2).~~~ q_{D_1+2d_1+1}+...+q_{d_2-1} &\leq x_{D_1+2d_1+1}+...+x_{d-1}\\
               \vdots & \\
               (d_1+k-2).~~~ q_{D_1+(k-2)d_1+1}+...+q_{d_2-1} &\leq x_{D_1+(k-2)d_1+1}+...+x_{d-1} \\
               (d_1+k-1).~~~ q_{D_1+(k-1)d_1+1}+...+q_{d_2-1} &\leq x_{D_1+(k-1)d_1+1}+...+x_{d-1}
               \end{aligned}
            \end{eqnarray}

       Like in the previous case, we recast these inequalities in the following matrix form:
       
       \begin{eqnarray}\label{ap23}
         \mathbf{Q2}=
         \begin{blockarray}{*{9}{c} l}
           \begin{block}{*{9}{>{$\footnotesize}c<{$}} l}
           \textcolor{green}{0}&\textcolor{red}{1}&~\textcolor{blue}{2} &$\cdots$ &  & \textcolor{cyan}{l'}
             & &$\cdots $  & \textcolor{magenta}{$d_{1}-1$}
              \\
           \end{block}
           \begin{block}{[*{9}{c}]>{$\footnotesize}l<{$}}
            \textcolor{green}{q_0} & \textcolor{red}{q_1} &~~~ \textcolor{blue}{q_3} &  & & \textcolor{cyan}{q_{\frac{l'(l'+1)}{2}}} & & & \textcolor{magenta}{q_{\frac{d_1(d_1-1)}{2}}} &  \\
            \textcolor{red}{q_2} & \textcolor{blue}{q_4} &&&&&&&\textcolor{brown}{q_{D_1+1=\frac{d_1(d_1+1)}{2}}}& $~\textcolor{brown}{d_1}$\\
            \textcolor{blue}{q_5} &&&&&&&&$\vdots$ &  \textcolor{gray}{$d_1+1$}\\ 
             $\vdots$ &&&\textcolor{cyan}{q_{\frac{l'(l'+1)}{2}+m'}}&&&&& $\vdots$ &  \\
              $\vdots$ &&&&&&&& \textcolor{orange}{q_{D_1+(k-1)d_1+1}} & \textcolor{orange}{$d_1-1+k$}  \\
               $\vdots$ &&&&&&&& $\vdots$ & \\
            $\vdots$ &&&&&&&&\textcolor{teal}{0}& $\textcolor{teal}{d_2-1}$  \\
            \textcolor{cyan}{q_{\frac{l(l+1)}{2}+l}} &&&&&&&& $\vdots$ & \\
            $\vdots$ &&&&\textcolor{orange}{q_{D_1+(k-1)d_1+j}}=\textcolor{orange}{q_{d_2-1}}&&&& $\vdots$ &  \\
            $\vdots$ &&&&&&&& $\vdots$ & \\
            \textcolor{magenta}{x_{D_1=\frac{d_1(d_1-1)}{2}+d_1-1}}&&&&&&& &0& ~$\textcolor{black}{2d_1-2}$\\
                        \textcolor{brown}{q_{D_1+d_1}}&&&&&&& & $\vdots$&\\
            $\vdots$&&&&&&& & $\vdots$&\\
             $\vdots$ &&&&&&&&0& ~$\textcolor{violet}{d_2-1+k'}$ \\
             \textcolor{orange}{0} &&&&&&&& $\vdots$ &  \\
                $\vdots$&   &&&&&& &&\\
                  \textcolor{teal}{0}& &0&&&&& &0&~$\textcolor{olive}{d_1+d_2-2}$\\
            \end{block}
         \end{blockarray}
         \end{eqnarray}
          Under the given value of $(d_1,d_2)$, the integers $(k,j)$ are uniquely determined by the following constraint 
          \begin{equation}\label{ap24}
          D_1+(k-1)d_1+j=d_2-1 ~~;~~ 1 \leq j \leq d_1.
          \end{equation}
          
          Here again we define the quantity $R$ taking the off diagonal element
          \begin{eqnarray}\label{ap25}
          \begin{aligned}
           R&=1.(q_1+q_2)+2.(q_3+q_4+q_5)+ \cdots + (d_1-1)[q_{\frac{d_1(d_1-1)}{2}}+\cdots + q_{D_1}]+d_1[q_{D_1+1}+ \cdots + q_{D_1+d_1}] \\ &+ \cdots +(d_1-2+k)[q_{D_1+(k-2)d_1+1}+\cdots+q_{D_1+(k-1)d_1}]+ \cdots+ (d_1-1+k)[q_{D_1+(k-1)d_1+1}+\cdots+q_{D_1+(k-1)d_1+j}]
           \end{aligned}
          \end{eqnarray}
          
            Summing the L.H.S. of the inequalities (\ref{ap22}) we get 
            \begin{eqnarray}\label{ap26}
            R= \sum\limits_{i=1}^{d_1-1}i\sum\limits_{k'=0}^{i}q_{\frac{i(i+1)}{2}+k'}+\sum\limits_{k'=1}^{k-1}(d_1-1+k')\sum\limits_{j'=1}^{d_1}q_{D_1+(k'-1)d_1+j'}+(d_1-1+k)\sum\limits_{j'=1}^{j}q_{D_1+(k-1)d_1+j'}
            \end{eqnarray}
                  and  summing the R.H.S. gives
                  \begin{eqnarray}\label{ap27}
                  \begin{aligned}
                  R'&=1.(x_1+x_2)+2.(x_3+x_4+x_5)+ \cdots + (d_1-1)[x_{\frac{d_1(d_1-1)}{2}}+\cdots + x_{D_1}]+d_1[q_{D_1+1}+ \cdots + x_{D_1+d_1}] \\
                  &+ \cdots 
                   +(d_1-2+k)[x_{D_1+(k-2)d_1+1}+\cdots+x_{D_1+(k-1)d_1}]+ (d_1-1+k)[x_{D_1+(k-1)d_1+1}+\cdots+x_{d-1}] \\
                  &=\sum\limits_{i=1}^{d_1-1}i\sum\limits_{k'=0}^{i}x_{\frac{i(i+1)}{2}+k'}+\sum\limits_{k'=1}^{k-1}(d_1-1+k')\sum\limits_{j'=1}^{d_1}x_{D_1+(k'-1)d_1+j'}+~~(d_1-1+k)\sum\limits_{j'=D_1+(k-1)d_1+1}^{d-1}x_{j'}
                   \end{aligned}
                  \end{eqnarray} 
                  
                  We can add and subtract $R$ in ergotropic gap equation (\ref{ap1}),
                  \begin{eqnarray}\label{ap28}
                  \Delta_{EG} =\sum\limits_{i=1}^{d_1-1}ip_{i} + \sum\limits_{i=1}^{d_2-1}iq_{i}-R+R - Z.
                  \end{eqnarray}
                  Since $R \leq R'$,
                  \begin{eqnarray}\label{ap29}
                  \leq \sum\limits_{i=1}^{d_1-1}ip_{i} + \sum\limits_{i=1}^{d_2-1}iq_{i}-R+R' - Z
                  \end{eqnarray}
                  
                  Substituting the value of $R$, $R'$ from above and $Z$ from Eq.(\ref{ap7}) we get
                  \begin{eqnarray}\label{ap30}
                  \begin{aligned}
                  &\leq \sum\limits_{i=1}^{d_1-1}ip_{i} + \sum\limits_{i=1}^{d_2-1}iq_{i} \\
                  &-[\sum\limits_{i=1}^{d_1-1}i\sum\limits_{k'=0}^{i}q_{\frac{i(i+1)}{2}+k'}+\sum\limits_{k'=1}^{k-1}(d_1-1+k')\sum\limits_{j'=1}^{d_1}q_{D_1+(k'-1)d_1+j'}+(d_1-1+k)\sum\limits_{j'=1}^{j}q_{D_1+(k-1)d_1+j'}]\\
                  &+[\sum\limits_{i=1}^{d_1-1}i\sum\limits_{k'=0}^{i}x_{\frac{i(i+1)}{2}+k'}+\sum\limits_{k'=1}^{k-1}(d_1-1+k')\sum\limits_{j'=1}^{d_1}x_{D_1+(k'-1)d_1+j'}+~~(d_1-1+k)\sum\limits_{j'=D_1+(k-1)d_1+1}^{d-1}x_{j'}] \\&-[\sum\limits_{i=1}^{d_1-1}i\sum\limits_{k'=0}^{i}x_{\frac{i(i+1)}{2}+k'}+ \sum\limits_{k'=1}^{d_2-d_1}(d_1-1+k')\sum\limits_{j'=1}^{d_1}x_{D_1+(k'-1)d_1+j'}+ \sum\limits_{i=1}^{d_1-1}(d_2-1+i)\sum\limits_{j'=1}^{d_1-i}x_{D_2+(i-1)d_1-\frac{i(i-1)}{2}+j'}].
                  \end{aligned}
                  \end{eqnarray}

                  The integers $(k,j)$ are determined by Eq. (\ref{ap24}). The constraint $(k-1) \leq (d_2-d_1)$ implies that the term in $2^{nd}$ parentheses should always be less than the term in $3^{rd}$,  the difference being $\delta$.
                  \begin{eqnarray}\label{ap31}
                  \begin{aligned}
                  &\leq \sum\limits_{i=1}^{d_1-1}ip_{i} + \sum\limits_{i=1}^{d_2-1}iq_{i}\\
                  &-[\sum\limits_{i=1}^{d_1-1}i\sum\limits_{k'=0}^{i}q_{\frac{i(i+1)}{2}+k'}+\sum\limits_{k'=1}^{k-1}(d_1-1+k')\sum\limits_{j'=1}^{d_1}q_{D_1+(k'-1)d_1+j'}+(d_1-1+k)\sum\limits_{j'=1}^{j}q_{D_1+(k-1)d_1+j'}]-\delta
                  \end{aligned}
                  \end{eqnarray}
                  Distributing the 2nd term through the 3rd, 4th and 5th term we obtain
                  \begin{eqnarray}\label{ap32}
                  \begin{aligned}
                  &\leq \sum\limits_{i=1}^{d_1-1}ip_{i} +\sum\limits_{i=1}^{d_1-1}\sum\limits_{k=0}^{i}[\frac{i(i+1)}{2}+k-i]q_{\frac{i(i+1)}{2}+k}+\sum\limits_{k'=1}^{k-1}\sum\limits_{j'=1}^{d_1}[(D_1+(k'-1)d_1+j')-(d_1-1+k')]q_{D_1+(k'-1)d_1+j}\\ &+\sum\limits_{j'=1}^{j}[D_1+(k-1)d_1+j'-(d_1-1+k)] q_{D_1+(k-1)d_1+j'}-\delta.
                  \end{aligned}
                  \end{eqnarray}

                 The frequency of $(p_i,q_i)$ increases with $i$. To find the bound on ergotropic gap for separable states we need to maximize the above equation, i.e.,
                 \begin{eqnarray}\label{ap33}
                  max(\Delta_{EG}) \leq max_{p_i}(1^{st}) + max_{q_i}(2^{nd}+3^{rd}+4^{th})-min(\delta) \nonumber
                 \end{eqnarray}

                  As $\{p_i\}$ and $\{q_i\}$ are defined in decreasing order and $min(\delta) = 0$, the best choice is
                            \begin{eqnarray}\label{ap34}
                            (p_1,p_2, \cdots, p_{d_1-1})\equiv (\frac{1}{d_1},\frac{1}{d_1}, \cdots,\frac{1}{d_1})\nonumber \\
                            (q_1,q_2, \cdots, q_{d_2-1})\equiv (\frac{1}{d_2},\frac{1}{d_2}, \cdots,\frac{1}{d_2})
                            \end{eqnarray}

      Substituting (\ref{ap34}) in (\ref{ap31}) and $\delta=0$, we get the bound on ergotropic gap defined by $M(d_1,d_2)$ 
        
      \begin{eqnarray}
      \begin{aligned}
      \Delta_{EG} \leq M(d_1,d_2) & = \frac{1}{d_1}\sum\limits_{i=1}^{d_1-1}i + \frac{1}{d_2}\sum\limits_{i=1}^{d_2-1}i - \frac{1}{d_2}\sum\limits_{i=1}^{d_1-1}i(i+1)-\sum\limits_{k'=1}^{k-1}\frac{(d_1+k'-1)d_1}{d_2}-\frac{d_1+k-1}{d_2}\sum\limits_{j'=1}^{j}j'\\
     & = \frac{d_1+d_2}{2}-1-\frac{d_1}{d_2}[\frac{d^2_1-1}{3}+(k-1)(d_1-1+\frac{k}{2})]-\frac{j(j+1)(d_1-1+k)}{2d_2}
     \end{aligned}
      \end{eqnarray}      
      This gives us the dimension dependent criterion.
      
                   So from the spectral and dimension dependent criterion we finally conclude that if a  $d_1\times d_2$ state is separable then its ergotropic gap would be bounded by
                   
                   \begin{equation}
                    \Delta_{EG} \leq min\{(Y-Z),M(d_1,d_2)\},
                    \end{equation}\\

                    where  $M(d_1,d_2)$ = \Bigg\{ \begin{tabular}{ccc}
                      $\frac{d_1-1}{2}+\frac{d_2-1}{2} - \frac{l}{d_2}[\frac{l^2-1}{3}+m+1]$ ~~~~~~~~~~~~~~~~~~~~~~~~~~~~~~~~~~~~~~~~~~~~if & $(d_2-1) \leq D_1$ \\
                      $\frac{d_1+d_2}{2}-1-\frac{d_1}{d_2}[\frac{d^2_1-1}{3}+(k-1)(d_1-1+\frac{k}{2})]- \frac{j(d_1-1+k)(j+1)}{2d_2}$ & if & $d_2-1 > D_1$
                      \end{tabular}\\
                    \end{widetext}
    
\end{document}